\documentclass[runningheads]{svmult}

\usepackage{makeidx}   
\usepackage{graphicx}  
\usepackage{subeqnar}  
\usepackage{multicol}  
\usepackage{physprbb}  
\makeindex             

\begin{document}
\title*{Search for the Companions of
Galactic SNe Ia}
\toctitle{Search for the Companions of 
\protect\newline Galactic SNe Ia}
%
%
\titlerunning{Search for the Companions}
%

\author{Pilar Ruiz--Lapuente\inst{1,2}, Fernando Comeron \inst{3},
 Stephen Smartt \inst{4},\\ Robert Kurucz \inst{5} Javier Mendez
 \inst{1,6},
 Ramon Canal \inst{1},\\ Alex Filippenko \inst{7} \& Ryan Chornock \inst{7}}

\authorrunning{Ruiz--Lapuente et al.}
%
%

\institute{Department of Astronomy, University of Barcelona 
\and Max--Planck Institut f\"ur Astrophysik, Garching
\and European Southern Observatory
\and Institute of Astronomy, Cambridge
\and Harvard--Smithsonian Center for Astrophysics
\and Isaac Newton Group, La Palma 
\and Department of Astronomy, University of California at Berkeley}

\maketitle              

\begin{abstract}

The central regions of the remnants of Galactic SNe Ia have been
examined for the presence of companion stars of the exploded 
supernovae. We present the results of this survey for the historical
SN 1572 and SN 1006. The spectra of the stars are modeled to obtain 
Teff, log g and the metallicity. Radial velocities are obtained with
an accuracy of 5--10 km s$^{-1}$. Implications for the nature of the
companion star in SNeIa follow. 

\end{abstract}

\section{Introduction}

Type Ia supernovae have long been recognized as close binary systems
where one of the stars, a carbon-oxygen WD (C+O WD), undergoes a
thermonuclear runaway after reaching the explosive conditions at its
center. While the remnants and
the ejecta left by the explosion have been studied in great detail,
 little is known about
companion stars. Motivated by the lack of definitive
constraints on the progenitors of SNe Ia, a programme was started to
find the moving companion of the exploding WD (Ruiz--Lapuente 1997;
Canal, Mendez \& Ruiz--Lapuente 2001).  

\bigskip

Several possibilities for the companion  have been
proposed. It could be a white dwarf, giant, subgiant, or a main sequence
star. Each of these possibilities and in particular the expected
peculiar velocities are discussed below. 

\bigskip

{\bf White Dwarf.} A double--degenerate (DD) system, where the mass
donor is a second WD, should not leave any companion. The companion is
destroyed in the mass--transfer process. 
Although first estimates of the efficiency in producing SNe Ia
disfavor this option, as reported in this conference (Napiwotzki et
al. 2002), more double degenerate systems able to merge in less than a
Hubble time are being discovered.


\bigskip

{\bf Subgiant or Giant.} Close binaries consisting of WDs with
subgiant or giant companions (also called Algol--like systems) can
produce the growth in mass of the WD until it reaches the explosive
condition (Whelan \& Iben 1973; Hachisu, Kato \& Nomoto 1996, 1999; 
Hachisu et al. 1999). The transfer takes place when the
subgiant or giant fills its Roche lobe due to its thermonuclear
evolution and pours material onto the WD. The material transferred to
the WD is H, and depending on the accretion rate the WD will either
grow in mass up to the Chandrasekhar limit or ignite a detonation in
an outer shell. Once the WD explodes, the ejecta hit the
companion. That produces several effects: the system is disrupted,
the companion moves with the orbital velocity it had before the
explosion, plus the kick velocity it acquires from the impact of the
ejecta. The interaction of the ejecta with the companion will also
strip away part of its mass. In addition, the bound remnant of the
companion will have its hydrostatic and thermal equilibrium altered
(Marietta, Burrows \& Fryxell 1999; Canal, M\'endez \& Ruiz--Lapuente
2001). Hydrostatic equilibrium will resume on a very short time scale,
but thermal equilibrium will only be regained on a longer, global 
Kelvin--Helmholtz time scale (see as well the earlier calculations
 done by Colgate 1970; Wheeler, Lecar \& McKee 1975; Fryxell \& Arnett 1981; 
Taam \& Fryxell 1985; Livne, Tuchman \& Wheeler 1992). That means the 
companion will exhibit an increase in radius, increased surface temperature, 
and an increased luminosity, which should be observable in the most recent 
Galactic Type Ia supernovae. Depending on the separation between the two 
stars at the time of explosion, the companion will acquire different 
velocities.  As a result of the orbital motion that the star had before the 
explosion and the kick velocity, we expect the star to be moving at 
velocities of a few hundred km s$^{-1}$ (see Table 1). That is one order of 
magnitude larger than the systemic velocities of typical stars at the same 
location in our Galaxy. The motion will be an identifying signature of the
companion.

\bigskip

{\bf Main Sequence Star.}  The final possibility for a SN progenitor
which has been
proposed consists of a WD plus a main--sequence companion, i.e., a
cataclysmic variable. Orbital shrinkage is driven in those systems by
magnetic braking plus the emission of gravitational wave radiation
(see Ruiz--Lapuente, Canal,
\& Burkert 1997, and references therein).  In this case, the moving
companion remaining after the explosion will be fainter than in 
the giant/subgiant case, but typical velocities will be higher.

\bigskip

Several considerations lead to the selection of SNIa remnants for the
detection of the companion star within this project. 
First, the X--ray shell morphology should be spherically symmetric
with a well defined center of the remnant. The distance and the age
of the SNeIa remnants should be in an adequate range to allow a
reasonable search radius (see Table 1). 
 SNe of Type Ia such as 
Tycho (SN 1572) and SN 1006, which have shell morphologies
 with high spherical
symmetry preserved up to 1000 yr after the
explosion seem to be the most natural targets for this study.  

\bigskip

\begin{table}
\begin{center}
\footnotesize\rm
\caption{Typical apparent magnitudes, proper motions, radial velocities, 
and maximum angular distance from explosion site (after 10$^{3}$ yr) of SNeIa 
companions}
\bigskip
\renewcommand{\arraystretch}{1.4}
\setlength\tabcolsep{5pt}
\begin{tabular}{lllll}
\hline\noalign{\smallskip}

Companion type  & m$_{V}$ & $\pi$ (arcsec yr$^{-1}$)  & v$_{r}$ (km s$^{-1}$)& 
 $\theta$ (arcmin)\\

and distance    & & & & \\

\hline

                & & & & \\
Main sequence   &      &      &     &          \\
1 kpc           & 15.1 & 0.067& 320 &  1.6     \\
5 kpc           & 18.6 & 0.013& 320 &  0.3     \\
                & & & & \\
Subgiant        &      &      &     &          \\
1 kpc           & 12.6 & 0.038& 180 &  0.9     \\
5 kpc           & 16.1 & 0.008& 180 &  0.2     \\
                & & & & \\
Red giant       &      &      &     &          \\
1 kpc           & 10.5 & 0.015& 70  &  0.4     \\
5 kpc           & 14.0 & 0.003& 70  &  0.1     \\

                & & & & \\
\hline
\end{tabular}
\end{center}
\end{table}

\bigskip

\section{SN 1572} 

SN 1572 (Tycho Brahe's supernova) 
is close to the Galactic plane ( ${\it b} =$ +1.4 $^{0}$).
 The field of the supernova is
3.8 $'$ in radius. 
 In SN 1572 the radio shell is very regular  and early
 radio and optical expansion
 measurements have found similar results on the ejecta  
 expansion rate. The distance inferred by the expansion of the radio shell
and by other methods lies between 2.25 and 4.5 kpc 
(Strom, Goss \& Shaver 1982). 

\bigskip

\noindent
 An estimate of the center is possible with an uncertainty less
than 10 $\%$ of the radio shell. Therefore, it
 seemed a good strategy to complete observations 
 down to a magnitude limit m$_{R}$ $\sim$  23 of 
 the stars within 0.7 $'$ of the center.  
 In Figure 1 we show the spectra of some of the stars near the center 
 of the remnant.  A red giant is very near the geometrical center of SN 1572. 
 Other stars in the vicinity range from supergiants to WDs. 

\bigskip

  We have obtained spectra with high enough resolution
  to allow detection of motions in the radial direction. Our
  spectra correspond to different epochs and this allows an additional
  check of variation of the velocities in those directions.   
  Spectra were obtained with ISIS and UES
  at the William Herschel Telescope
  and with ESI at the Keck Telescope (see Table
  2 for the observations). 

\bigskip

\begin{table}
\caption{Observations of SN 1572}
\begin{center}
\renewcommand{\arraystretch}{1.4}
\setlength\tabcolsep{5pt}
\begin{tabular}{lllllll}
\hline\noalign{\smallskip}
Run  & Rd ($'$)$^{1}$  & m$_{R}$ $^{2}$  & Telescope $^{3}$  & R
  &  Spec Range (A) & stellar types  \\
\noalign{\smallskip}
\hline
\noalign{\smallskip}
 (1) & 0.7 & 14 & WHT (UES) & 50,000 & 4000--7100 & red giant \\
 (2) & 0.7 & 23 & WHT (ISIS) & 15,000  & 4600-7500  & red giants to WD \\
 (3) & 0.7 & 23 & Keck (ESI) & 7000 & 4000--10000  & as above \\
\hline
\end{tabular}
\end{center}
\label{Tab1a}

\noindent$^{1}$ Radius of the search

\noindent$^{2}$ Limiting magnitude 

\noindent$^{3}$ Telescopes(Instrumentation) 

\end{table}

\begin{table}
\caption{Observations of SN 1006}
\begin{center}
\renewcommand{\arraystretch}{1.4}
\setlength\tabcolsep{5pt}
\begin{tabular}{lllllll}
\hline\noalign{\smallskip}
Run  & Rd ($'$)$^{1}$  &  m$_{R}$ $^{2}$& Telescope $^{3}$ & R
  & Spec range (A) & stellar types \\
\noalign{\smallskip}
\hline
\noalign{\smallskip}
 (1) & 5 & 13 & NTT (EMMI) & 10,000 & 3950--7660 & red giants \\
 (2) & 5 & 15 & VLT (UVES) & 50,000 & 3500--9000  &  all types \\
\hline
\end{tabular}
\end{center}
\label{Tab1a}
\noindent$^{1}$ Radius of the search

\noindent$^{2}$ Limiting magnitude 

\noindent$^{3}$ Telescopes(Instrumentation) 

\end{table}

\section{SN 1006} 

SN 1006 is at a Galactic latitude {\it b}= $+$ 14.6 $^{0}$,
about $\sim$ 550 pc 
above the Galactic plane. The field of
this supernova extends  15$'$ in radius. 
An examination of all the centers given in the literature up to now
(Winkler \& Long 1997; Reynolds and Gilmore 1986; van den Bergh 1976) and our inspection
of the geometry of the SNIa, suggest that within 1$'$, the center 
of SN 1006 should be at $\alpha$= 15 02 55, $\delta$=-41 55 12 (J2000) as
given by Allen et al. (2001). 
 The distance estimates to this SNR are in the range between  
1.5 and 2.5 kpc. Recently Winkler, Gupta \& Long (2002) 
measure a distance of 2.17 $\pm$ 0.08 kpc to SN 1006 from the 
expansion rate of the remnant as derived from the optical filaments.
Given the predictions for the movement of the 
companion, at a distance between 1.5 and 2.5 kpc, 
a search  around 5$'$ at the best
determined center of the remnant should find the companion of this 
SNIa.  

\smallskip

Our search goes down to mag 15 in {\it R}. At the distance to this remnant,
this means to reach stars of solar luminosity. Thus, an 
exhaustive test of the hypothesis of supergiant, giant and main 
sequence stars  is obtained in this way (see Table 3 for a summary 
of the observations).

\begin{figure}[hbtp]
\begin{center}
\includegraphics[width=1\textwidth]{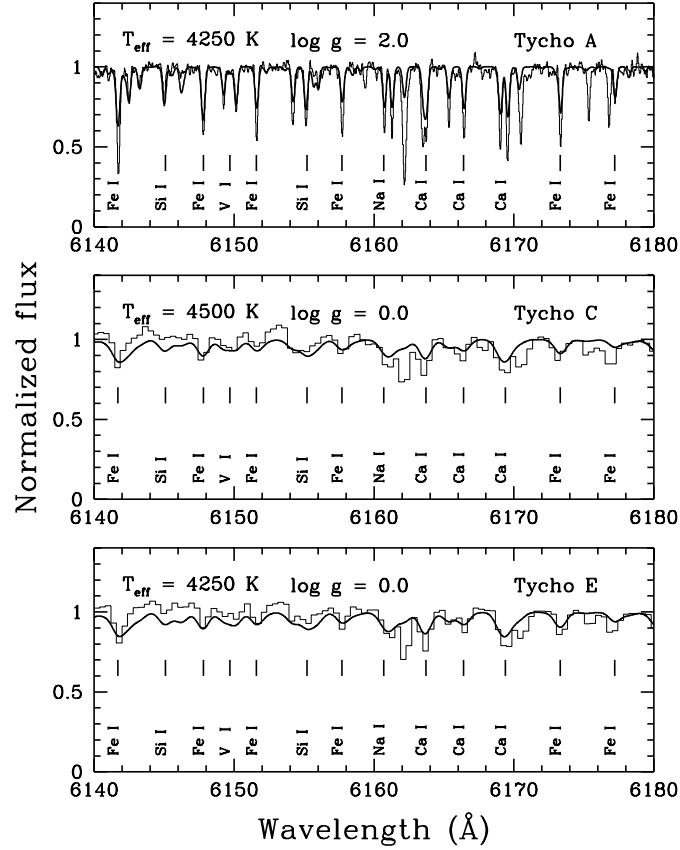}
\end{center}
\caption[]{ 
Calculated synthetic spectra compared with the observed spectra of
the SN companion candidates near the center of SN 1572.
These are the closest red giants and 
supergiants to the center of the explosion, their surface gravity goes
from log g=2 for Tycho A to log g=0 for C and E. The effective 
temperatures for those stars are similar. They
are in the range 4200--4500 K. Model atmospheres with solar chemical 
abundances give a good account of the spectra. 
The overall spectral comparison allow us to
exclude overabundances of 
the Fe--peak elements. Moreover, the stars show no enhancement of 
iron--peak elements versus intermediate--mass elements in the
spectra. Synthetic spectra are shown with bold continuous lines. 
}
\label{fig1}
\end{figure}

\begin{figure}[hbtp]
\begin{center}
\includegraphics[width=1\textwidth]{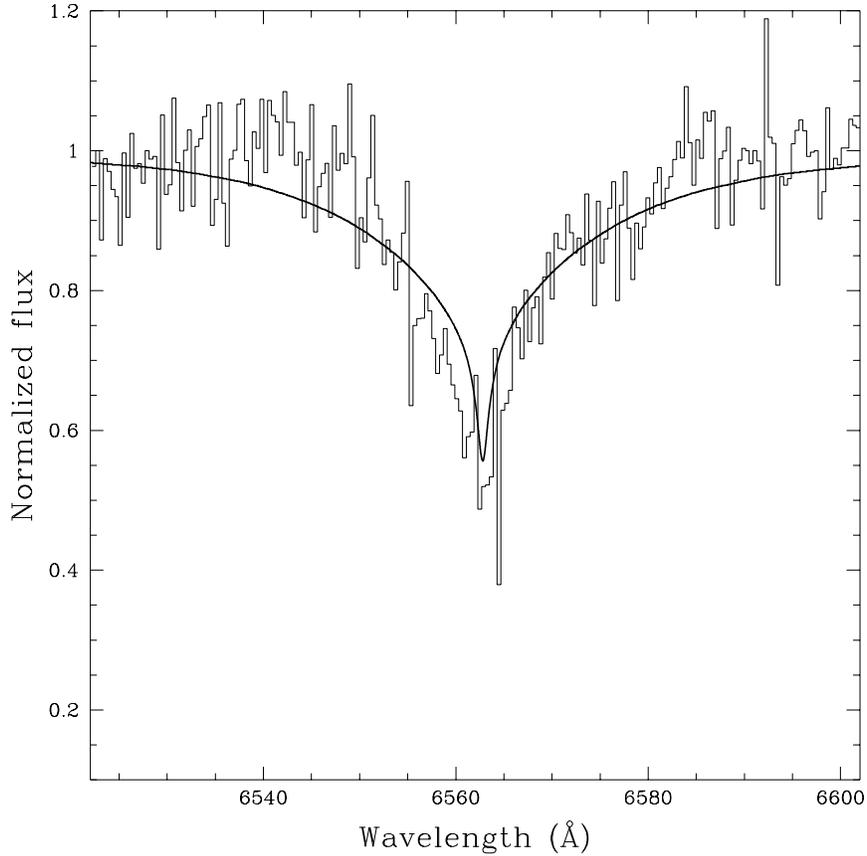}
\end{center}
\caption[]{ 
DA WD near the center of Tycho remnant. A preliminar analysis 
suggests log g=6, and 
Teff=25,000 K.
}
\label{fig1}
\end{figure}

\section{Modeling}

We have compared calculated LTE spectra with the observed spectra of
the SN companion candidates. We have used
Kurucz's grids of model atmospheres (Kurucz 1993), a spectral synthesis code 
based on the Uppsala Synthetic Spectrum Package (Gustafsson et al. 1975),
 and the atomic data 
from the linelists in Kurucz. The profiles of the Balmer lines of H, 
calculated separately for the model atmospheres of the grids, have also 
been used in the spectral fits. The comparison of the 
synthetic spectra with the observations is shown in Fig. 1 for some
of the stars 
in our sample of candidates to SN companion. Identifications of the most 
significant lines are given. Good agreement is achieved assuming solar 
abundances (Anders \& Grevesse 1989). Moreover, 
none of the closest stars to the center show signs
 of any spectroscopic anomaly. Radial 
velocities have also been measured from the wavelength
 shifts of several lines in each 
observed spectrum: they are in the range  --40 to --60 km 
s$^{-1}$, which is perfectly attribuable to Galactic rotation alone.

\bigskip

The low--velocity tail of the ejecta of a thermonuclear (Type Ia) 
supernova is made of Fe--peak elements (Fe, Co, Ni, Mn, Cr, V, Ti), 
as it comes from the material at the center of the exploding star and it thus 
reached the highest temperatures ($\sim$ 10$^{10}$ K), being then processed 
to Nuclear Statistical Equilibrium (Thielemann 1989).
 Since this material is the most 
likely to have contaminated the surface of a binary companion (see
papers cited in section 1), especial care 
has been given to the determination of the abundances of those elements in 
the atmospheres of the stars in our sample. An overabundance of those
elements in relation to intermediate--mass elements is expected.

\bigskip

  The giants at the core do not show any signs of contamination and
  are consistent with solar metallicities (or slightly lower
  solar metallicities). 
  This result disfavors the subgiant possibility and that
  of a red giant closely bound to the WD prior to the moment of 
  explosion. 

\bigskip

  Strong constraints can be placed also on the main sequence star 
 candidates.
  No main sequence star moving at high radial velocity has been 
  detected in SN 1572. An examination over a more
  extended radius around the center
  is being done.
 
\bigskip

  For SN 1006 we have found radial velocities in the range 30--120 km 
 s$^{-1}$ for the giant stars within the radius of search. Stars show
 a larger dispersion in radial velocity than in the field of the
 remnant of SN 1572. 
 A full account of the observations, the measurement of the reddening, 
 the overall spectral modeling and a discussion on the distance
 to those candidates will be presented elsewhere.

\bigskip

  This paper is based on observations obtained at the WHT, 
  operated by the Isaac Newton Group of Telescopes, funded by
  the PPARC at the Spanish Observatorio del Roque de los 
  Muchachos of the Instituto de Astrofisica de Canarias; at
  W.M. Keck Observatory, which is operated as a scientific 
  partnership among the California Institute of Technology, 
  the University of California and the National Aeronautics
  and Space Administration (the Observatory was made possible 
  by the generous financial support of the W.M.Keck Foundation); 
  and the NTT and VLT at the La Silla and Paranal Observatories
  of ESO for programs 67.D-0348(A) and 69.D-0397(A). 
  We express our gratitude to the ESO User Support Group for providing
  help and conducting service observations and pipeline reductions of
  the data taken at Paranal. P.R.L would like to thank as well
  the assistance provided by Carlos Abia with the Uppsala Synthetic
  Spectrum Package. 
  
\vfill\eject

\end{document}